# A Note to Article "Quantization of Light Energy Directly from Classical Electromagnetic Theory in Vacuum"

Wei-Long She

Institute for Lasers & Spectroscopy, Sun Yat-Sen University, Guangzhou, China

Very recently we present a theory [1] to show that the quantization of light energy in vacuum can be derived directly from classical electromagnetic theory through the consideration of statistics based on classical physics and reveal that the quantization of energy is an intrinsic property of light as a classical electromagnetic wave and has no need of being related to particles. In this theory a key concept of stability of statistical distribution was involved. Here is a note to the theory, which would be helpful for understanding the concept of stability of statistical distribution.

Suppose $\rho(q)$ be the density function of the generalized coordinate of 1-D harmonic oscillator (ODHO), $q$, to be found, then we can always introduce such a complex-valued function $\psi(q)$ mathematically that makes $|\psi(q)|^2 = \rho(q) \geq 0$. The $\psi(q)$ should then satisfy the following universal conditions: i) $\int_{-\infty}^{\infty} |\psi(q)|^2 \, dq = 1$; ii) $\lim_{|q|\to\infty} \psi(q) = 0$; iii) $\int_{-\infty}^{\infty} 1/2 \cdot \omega^2 q^2 |\psi(q)|^2 \, dq \leq E < \infty$. These are understandable ones in classical physics: i) is the general condition for any density function; ii) ensures the amplitude of each ODHO being finite; and iii) is the finiteness condition of the expectation value, $E$, of the ODHO's energy. One can see that those meeting these conditions make up of the whole of the function candidates used to construct $\rho(q)$ and the number of them is infinite. To find those of relatively stable statistical distributions of $q$, we should change i) - iii) into a set of equivalent "equilibrium conditions"(EECs)[1] (see below). To make the concept of stability of statistical distribution be understood easily, let us draw an analogy as follows:

$1/2 \int_{-\infty}^{\infty} |\frac{d\psi(q)}{dq}|^2 \, dq - 1/2 \int_{-\infty}^{\infty} \omega^2 L^2 |F(L)|^2 \, dL = 0$,

$1/2 \int_{-\infty}^{\infty} |\frac{dF(L)}{dL}|^2 \, dL - 1/2 \int_{-\infty}^{\infty} \omega^2 q^2 |\psi(q)|^2 \, dq = 0$,

$\int_{-\infty}^{\infty} |F(L)|^2 \, dL - 1 = 0$,

$\int_{-\infty}^{\infty} |\psi(q)|^2 \, dq - 1 = 0$,

$\lim_{|q|\to\infty} \psi(q) = 0$,

$\lim_{|q|\to\infty} \int_{-\infty}^{\infty} F(L) \exp(-i\omega qL) \, dL = 0$.

**The ideal law $\psi(q)$ and $F(L)$ obey (EECs, see Ref. [1])**

**Analogue**

$x^2 - y^2 = 0$

**The ideal law $x$ and $y$**

The real statistical distributions of $q$ cannot be just $|\psi(q)|^2$ coming from the rigid solution of EECs [1]. It will encounter some unpredictable but ineluctable tiny perturbations, $\delta\psi$, independent of the solutions of EECs, i.e., it should be $|\psi(q) + \delta\psi|^2$ but unpredictable (see Fig.1 below). All the rigid solution $\psi(q)$ can be divided into two types. One is scarcely different from $\psi(q) + \delta\psi$, i.e., it is also the optimized solution of overdetermined equations

$1/2 \int_{-\infty}^{\infty} |\frac{d[\psi(q) + \delta\psi_i]}{dq}|^2 \, dq - 1/2 \int_{-\infty}^{\infty} \omega^2 L^2 |F(L) + \delta F_i|^2 \, dL = 0$,

$1/2 \int_{-\infty}^{\infty} |\frac{d[F(L) + \delta F_i]}{dL}|^2 \, dL - 1/2 \int_{-\infty}^{\infty} \omega^2 q^2 |\psi(q) + \delta\psi_i|^2 \, dq = 0$,

$\int_{-\infty}^{\infty} |F(L) + \delta F_i|^2 \, dL - 1 = 0$,

$\int_{-\infty}^{\infty} |[\psi(q) + \delta\psi_i|^2 \, dq - 1 = 0$,   (**i = 1,2,3,..., n; n >> 2**)

The really physical quantities cannot be just those coming from the rigid solutions of equation $x^2 - y^2 = 0$. They will encounter some unpredictable but ineluctable tiny perturbations, $\delta x$ and $\delta y$, independent of the solutions of the equation, i.e., they should be ($x + \delta x$, $y + \delta y$) but unpredictable. All the rigid solution ($x$, $y$) can be divided into two types. One is scarcely different from ($x + \delta x$, $y + \delta y$), i.e. it is also the optimized solution of overdetermined equations

$(x + \delta x_1)^2 - (y + \delta y_1)^2 = 0$,

$(x + \delta x_2)^2 - (y + \delta y_2)^2 = 0$,

$(x + \delta x_3)^2 - (y + \delta y_3)^2 = 0$,

$$\lim_{|q|\to\infty}[[\psi(q)+\delta\psi_i]=0,$$

$$\lim_{|q|\to\infty}\int_{-\infty}^{\infty}[F(L)+\delta F_i]\exp(-i\omega qL)dL=0$$

(where $\delta\psi_1,\delta\psi_2,...\delta\psi_n$ stand for the possible forms of $\delta\psi$ and $\delta F_1,\delta F_2,...\delta F_n$ stand for those of $\delta F$ introduced on the consideration of symmetry, which is the tiny perturbations on $F(L)$ independent of the solutions of EECs and therefore independent of $\delta\psi$ and can be zero), and another is not. The first one resists the perturbations and is relatively stable while the second is too sensitive to the perturbations therefore very unstable that it cannot hold itself. So the second cannot exist really in physics.

**Question**: which $(\psi(q), F(L))$ is the optimized solution of the above equations?

**Answer:** should be those satisfying the following equations [1].

$-1/2\cdot d^2/dq^2\cdot\psi(q)+(-\lambda_1)1/2\cdot\omega^2 q^2\psi(q)+\lambda_3\psi(q)=0$,

$-\lambda_1/2\cdot d^2/dL^2\cdot F(L)-1/2\cdot\omega^2 L^2 F(L)+\lambda_2 F(L)=0$.

where $\lambda_i$ ($i=1,2,3$) denote three Lagrange's multipliers.

**The answer can be gotten by introducing functional with two unknown functions**

$I(\psi(q),F(L))=1/2\int_{-\infty}^{\infty}|\frac{d\psi(q)}{dq}|^2\,dq-1/2\int_{-\infty}^{\infty}\omega^2 L^2\,|F(L)|^2\,dL$,

$1/2\int_{-\infty}^{\infty}|\frac{dF(L)}{dL}|^2\,dL-1/2\int_{-\infty}^{\infty}\omega^2 q^2\,|\psi(q)|^2\,dq=0$,

$\int_{-\infty}^{\infty}|F(L)|^2\,dL-1=0$,

$\int_{-\infty}^{\infty}|\psi(q)|^2\,dq-1=0$,

$\lim_{|q|\to\infty}\psi(q)=0$,

$\lim_{|q|\to\infty}\int_{-\infty}^{\infty}F(L)\exp(-i\omega qL)dL=0$

and performing $\delta I(\psi(q),F(L))=0$ as did Ref.[1].

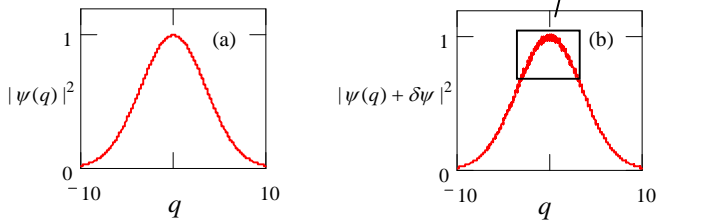

Fig.1 the comparison between ideal and real statistical distributions of $q$: (a) ideal; (b) real.

……………………………

$(x+\delta x_n)^2-(y+\delta y_n)^2=0\quad(n\gg 2)$

(where $\delta x_1,\delta x_2,...\delta x_n$ stand for the values of $\delta x$ possible and $\delta y_1,\delta y_2,...\delta y_n$ stand for those of $\delta y$, simulated by some of random numbers), and another is not. The first one resists the perturbations and is relatively stable while the second is too sensitive to the perturbations therefore very unstable that it cannot hold itself (see Fig.2 below).

**Question:** which $(x,y)$ is the optimized solution of the above equations?
**Answer:** should be (0,0), which can be approached by using optimization method.

The idea described above comes from classical physics especially classical experimental physics.

**The answer can also be gotten by introducing function**

$$f(x,y)=x^2-y^2$$

and performing

$\partial f/\partial x=0$,
$\partial f/\partial y=0$.

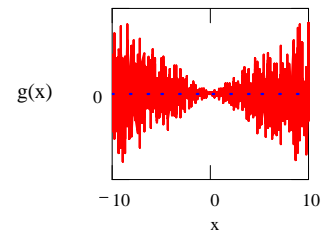

Fig.2 the stability of the solutions of equation $x^2-y^2=0$, Where $\delta x$ and $\delta y$ are two random numbers; $g(x)=(x+\delta x)^2-(x+\delta y)^2$ for $y=x$. Similarly when $y=-x$.